\documentclass [12 pt]{article}
\def\be{\begin{equation}}
\def\eea{\end{eqnarray}}
\def\bea{\begin{eqnarray}}
\def\ee{\end{equation}}

\usepackage[psamsfonts]{amssymb}
\usepackage{amsmath}
\author{F. Kheirandish$^{1}$ \footnote{fardin$_{-}$kh@phys.ui.ac.ir} and M.
Amooshahi$^{1}$ \footnote{amooshahi@sci.ui.ac.ir}
\\ $^{1}$ {\small Department of Physics, University of Isfahan,}
\\ {\small Hezar Jarib Ave., Isfahan, Iran.}}
\title{Spin Precession and Quantum Vacuum}
\begin{document}
\maketitle
\begin{abstract}
\noindent \\
The effect of quantum vacuum on spin precession is investigated.
The radiation reaction is obtained and the time of spin flip
 (up state to down state) or spontaneous decay, is calculated.\\\\
 \end{abstract}
\section{Introduction}
In QED, a charged particle in quantum vacuum interacts with the
vacuum field and its own field known as radiation reaction. In
classical electrodynamics, there is only the radiation reaction
field that acts on a charged particle in the vacuum. The vacuum
and radiation reaction fields have a fluctuation-dissipation
connection and both are required for the consistency of QED. For
example the stability of the ground state, atomic transitions and
lamb shift can only be explained by taking into account both
fields. If self reaction was alone the atomic ground state would
not be stable [1]. When a quantum mechanical system interacts
with the  vacuum quantum field, the coupled Heisenberg equations
for both system and field give us the radiation reaction field,
for example it can be shown that the radiation reaction for a
charged harmonic oscillator is $ \frac{2e^2}{3c^3} $ [2].\\
Dirac$^,$s theory of emission and absorption [3] was the first
application of the quantum theory of radiation. Dirac argued that
his theory must give the effect of radiation reaction on the
emitting systems. Spantaneous emission was also interpreted in
terms of radiation reaction in the theory of Landau  and before
the development of quantum formalism by Van Vlec [4]. However
contemporary physicits generally invoke vacuum electromagnetic
field or fluctuating field give a physical explanation for the
occurrence of spontaneous emission [5]. These two interpretations
based on radiation reaction or vacuum field fluctuations are
infact closely related in the quantum theory of radiation [2].
Spantaneous emission can be correctly described only when the
radiation field is quantized, if the field is not treated quantum
mechanically we obtain predictions in conflict with experiment. In
this paper we obtain in section 2, radiation reaction magnetic
field of a spin-$\frac{1}{2}$ system and then
 in section 3, we will show how the effect of quantum vacuum field
  cause a spin flip or a spontaneously decay from up state to down
  state.
\section{Radiation reaction}
Let us assume a magnetic field $ \vec{B}_L=B_L\hat{z} $ in
direction of $ z $ axis in laboratory exerted on a spin-$
\frac{1}{2}$ particle . The Hamiltonian that specify the time
evolution of the components of the spin in the absence of the
magnetic field of the quantum vacuum, is
\begin{equation}\label{S1}
H_0=\omega S_z\hspace{2.00cm}\omega=\frac{|e|B_L}{m}
\end{equation}
where $ e $ and $ m $ are  the charge and the mass of the
particle. In the following we will show how the presence of the
magnetic field of the quantum vacuum leads to a spontaneous decay
or a spin flip.\\
The vector potential $ \vec{A} $ and canonical momentum density $
\vec{\pi}_F $  of the quantum vacuum field in coulomb gauge are
\begin{eqnarray}\label{S2}
&&\vec{A}(\vec{r},t)=\sum_{\lambda=1}^2\int
d^3\vec{k}\sqrt{\frac{\hbar}{2(2\pi)^3\varepsilon_0\omega_{\vec{k}}}}
[a_{\vec{k}\lambda}(t)e^{i\vec{k}.\vec{r}}+a_{\vec{k}\lambda}^\dag(t)
e^{-i\vec{k}.\vec{r}}]\vec{e}(\vec{k},\lambda)\nonumber\\,
&&\vec{\pi}_F(\vec{r},t)=-i\sum_{\lambda=1}^2\int
d^3\vec{k}\sqrt{\frac{\hbar\omega_{\vec{k}}}{2(2\pi)^3}}
[a_{\vec{k}\lambda}(t)e^{i\vec{k}.\vec{r}}-a_{\vec{k}\lambda}^\dag(t)e^{-i\vec{k}.\vec{r}}]
\vec{e}(\vec{k},\lambda).\nonumber\\
&&
\end{eqnarray}
where $\omega_{\vec{k}}=c|\vec{k}|$ and $\vec{e}(\vec{k},\lambda)$
are polarization unit vectors
\begin{eqnarray}\label{S3}
&&\vec{e}(\vec{k},\lambda).\vec{e}(\vec{k},\lambda')=\delta_{\lambda\lambda'},\nonumber\\
&&\vec{k}.\vec{e}(\vec{k},\lambda)=0,\hspace{1.50 cm}\lambda=1,2.
\end{eqnarray}
The creation and annihilation operators $
a_{\vec{k}\lambda}^\dag$, $ a_{\vec{k}\lambda}$, in any instant of
time satisfy the commutation relations
\begin{equation}\label{S4}
[a_{\vec{k}\lambda}(t),a_{\vec{k'}\lambda'}^\dag(t)]=\delta_{\lambda\lambda'}
\delta(\vec{k}-\vec{k'}),
\end{equation}
and their time dependence is to be determined from the total
Hamiltonian. Commutation relations (\ref{S4}) lead to
\begin{equation}\label{S5}
[\vec{A}_i( \vec{r},t),\pi_{Fj}(
\vec{r'},t)]=i\hbar\delta_{ij}^\bot(\vec{r}-\vec{r'}),
\end{equation}
where $\delta_{ij}^\bot(\vec{r}-\vec{r'})=\frac{1}{(2\pi)^3}\int
d^3\vec{k}e^{i\vec{k}.(\vec{r}-\vec{r'})}(\delta_{ij}-\frac{k_ik_j}{|\vec{k}|^2})
$, is the transverse delta function. \\
When the spin-$\frac{1}{2}$ system interacts with the laboratory
magnetic field $\vec{B}_L$ and the quantum vacuum, the total
Hamiltonian can be written as
\begin{equation}\label{S6}
H=\omega
S_z+\alpha\vec{S}.\vec{B}+\sum_{\lambda=1}^2[a_{\vec{k}\lambda}^\dag(t)a_{\vec{k}\lambda}(t)
+\frac{1}{2}],
\end{equation}
where $ \alpha=\frac{|e|}{m}$ and
\begin{equation}\label{S7}
\vec{B}=\nabla\times\vec{A}=\sum_{\lambda=1}^2\int
d^3\vec{k}\sqrt{\frac{\hbar}{2(2\pi)^3\omega_{\vec{k}}}}(i\vec{k}\times
\vec{e}_{\vec{k}\lambda})
[a_{\vec{k}\lambda}(t)-a_{\vec{k}\lambda}^\dag(t) ],
\end{equation}
is the magnetic field of the vacuum quantum field in the place of
the system which is taken here $\vec{r}=0$.\\
By using  (\ref{S4}) it is easy to show that the Heisenberg
equation for $a_{\vec{k}\lambda}$ is
\begin{equation}\label{S8}
\dot{a}_{\vec{k}\lambda}=\frac{i}{\hbar}[H,a_{\vec{k}\lambda}]=-i\omega_{\vec{k}}
a_{\vec{k}\lambda}
-\sqrt{\frac{\alpha^2}{2(2\pi)^3\hbar\omega_{\vec{k}}}}\vec{S}\cdot(\vec{k}
\times\vec{e}_{\vec{k}\lambda}).
\end{equation}
One can solve this equation formally
\begin{equation}\label{S9}
a_{\vec{k}\lambda}(t)=a_{\vec{k}\lambda}(0)e^{-i\omega_{\vec{k}}t}-
\sqrt{\frac{\alpha^2}{2(2\pi)^3\hbar\omega_{\vec{k}}}}
(\vec{k}\times\vec{e}_{\vec{k}\lambda})\cdot\int_0^t d
t'\vec{S}(t')e^{-i\omega_{\vec{k}}(t-t')}.
\end{equation}
Substitution of $a_{\vec{k}\lambda}(t)$ from (\ref{S9}) in
(\ref{S7}) gives two magnetic fields, the vacuum magnetic field
\begin{equation}\label{S10}
\vec{B}_0=\sum_{\lambda=1}^2\int
d^3\vec{k}\sqrt{\frac{\hbar}{2(2\pi)^3\omega_{\vec{k}}}}(i\vec{k}\times
\vec{e}_{\vec{k}\lambda})
[a_{\vec{k}\lambda}(0)e^{-i\omega_{\vec{k}}t}-a_{\vec{k}\lambda}^\dag(0)e^{i\omega_{\vec{k}}t}],
\end{equation}
and the radiation reaction magnetic field
\begin{equation}\label{S11}
\vec{B}_{RR}=-\frac{\alpha}{3\pi^2c^5}\int_0^\infty
d\omega_{\vec{k}}\omega_{\vec{k}}^3\int_0^t
dt'\vec{S}(t')\sin\omega_{\vec{k}}(t-t')=
-\frac{\alpha}{3\pi^2c^5}\frac{\partial^3\vec{S}(t)}{\partial
t^3}.
\end{equation}
\section{Spontaneous decay}
Spontaneous decay is responsible for most of the light around us.
For a thermal source the ratio of the spontaneous and simulated
emission rates for radiation of frequency $\omega_0$ is
$e^{\frac{\hbar\omega_0}{KT}}-1$. As mentioned in the
introduction, Spontaneous emission is a concequence of vacuum
fluctuating field or radiation reaction [4],[5]. In this section
we use this theory for calculatiing the spontaneous decay of a
spin-$\frac{1}{2}$ system interacting with the
magnetic field of the quantum vacuum field.\\
Using the commutation relations
\begin{equation}\label{S12}
[S_l,S_j]=i\hbar\varepsilon_{ljm}S_m,
\end{equation}
one can easily obtain the Heisenberg equations for spin components
\begin{eqnarray}\label{S13}
&&\dot{S}_x=-\omega
S_y+\alpha\sum_{\lambda=1}^2\int d^3\vec{k}\sqrt{\frac{\hbar}{2(2\pi)^3\omega_{\vec{k}}}}(i\vec{k}\times\vec{e}_{\vec{k}\lambda})_y[S_z(t)a_{\vec{k}\lambda}(t)-a_{\vec{k}\lambda}^\dag (t)S_z(t))]\nonumber\\
&&-\alpha\sum_{\lambda=1}^2\int d^3\vec{k}\sqrt{\frac{\hbar}{2(2\pi)^3\omega_{\vec{k}}}}(i\vec{k}\times\vec{e}_{\vec{k}\lambda})_z[S_y(t)a_{\vec{k}\lambda}(t)-a_{\vec{k}\lambda}^\dag (t)S_y(t))]\nonumber\\
&&\dot{S}_y=\omega S_x+\alpha\sum_{\lambda=1}^2\int d^3\vec{k}\sqrt{\frac{\hbar}{2(2\pi)^3\omega_{\vec{k}}}}(i\vec{k}\times\vec{e}_{\vec{k}\lambda})_z[S_x(t)a_{\vec{k}\lambda}(t)-a_{\vec{k}\lambda}^\dag (t)S_x(t))]\nonumber\\
&&-\alpha\sum_{\lambda=1}^2\int
d^3\vec{k}\sqrt{\frac{\hbar}{2(2\pi)^3\omega_{\vec{k}}}}
(i\vec{k}\times\vec{e}_{\vec{k}\lambda})_x[S_z(t)a_{\vec{k}\lambda}(t)-a_{\vec{k},
\lambda}^\dag (t)S_z(t))]\nonumber\\
&&
\end{eqnarray}
\begin{eqnarray}\label{S13.5}
&&\dot{S}_z=\alpha\sum_{\lambda=1}^2\int d^3\vec{k}\sqrt{\frac{\hbar}{2(2\pi)^3\omega_{\vec{k}}}}(i\vec{k}\times\vec{e}_{\vec{k}\lambda})_x[S_y(t)a_{\vec{k}\lambda}(t)-a_{\vec{k}\lambda}^\dag (t)S_y(t))]\nonumber\\
&&-\alpha\sum_{\lambda=1}^2\int d^3\vec{k}\sqrt{\frac{\hbar}{2(2\pi)^3
\omega_{\vec{k}}}}(i\vec{k}\times\vec{e}_{\vec{k}\lambda})_y[S_x(t)
a_{\vec{k}\lambda}(t)-a_{\vec{k}\lambda}^\dag (t)S_x(t))].\nonumber\\
&&
\end{eqnarray}
Since equal-time spin and field operators commute, we can write
the Heisenberg equations (\ref{S13}) and (\ref{S13.5}) in
different but equivalent ways. For instance we can use the normal
ordering in which photon annihilation operators $
a_{\vec{k}\lambda} $ appear at the right and creation operators
$a_{\vec{k}\lambda}^\dag $ appear at the left hand of spin
operators,or we can use the antinormal ordering in which photon
annihilation operators $ a_{\vec{k}\lambda} $ appear at the left
and creation operators $a_{\vec{k}\lambda}^\dag $ appear at the
right hand of spin operators. In the following we
will use the normal ordering.\\
Let us assume that the spin-field coupling is sufficiently weak
so that we can approximate
\begin{eqnarray}\label{S14}
&&S_+(t')=S_x(t')+iS_y(t')\simeq S_+(0)e^{i\omega
t'}=S_+(0)e^{i\omega t}e^{i\omega (t'-t)}\simeq S_+(t)e^{i\omega
(t'-t)}\nonumber\\
&&S_-(t')=S_x(t')-iS_y(t')\simeq S_-(0)e^{-i\omega
t'}=S_-(0)e^{-i\omega t}e^{-i\omega (t'-t)}\simeq
S_-(t)e^{-i\omega
(t'-t)}\nonumber\\
&&S_z(t')\simeq S_z(t)
\end{eqnarray}
in the integrand of (\ref{S9}).This is called Markovian
approximation[6]. Now for integrals over $ t' $ in (\ref{S13}) and
(\ref{S13.5}) we have
\begin{eqnarray}\label{S15}
&&\int_0^t d
t'e^{i(\omega_{\vec{k}}-\omega)(t'-t)}=
-i[\frac{1-\cos(\omega_{\vec{k}}-\omega)t}{\omega_{\vec{k}}-\omega}]+
\frac{\sin(\omega_{\vec{k}}-\omega)t}{\omega_{\vec{k}}-\omega},\nonumber\\
&&\int_0^t d
t'e^{i(\omega_{\vec{k}}+\omega)(t'-t)}=-i[\frac{1-\cos(\omega_{\vec{k}}+
\omega)t}{\omega_{\vec{k}}+\omega}]+\frac{\sin(\omega_{\vec{k}}+\omega)t}
{\omega_{\vec{k}}+\omega},\nonumber\\
&&\int_0^t d
t'e^{i(\omega_{\vec{k}})(t'-t)}=-i[\frac{1-\cos\omega_{\vec{k}}t}{\omega_{\vec{k}}}]+
\frac{\sin\omega_{\vec{k}}t}{\omega_{\vec{k}}}.
\end{eqnarray}
In the first integral, the term inside the bracket, vanishes if $
\omega_{\vec{k}}-\omega=0 $, otherwise it is effectively $
\frac{1}{\omega_{\vec{k}}-\omega}$, because of the rapid
oscillations of $\cos(\omega_{\vec{k}}-\omega)t$ for $
(\omega_{\vec{k}}-\omega)t>>1 $. The second term in the first
integral effectively vanishes unless $\omega_{\vec{k}}-\omega=0
$, which  tends to $t$ in large times. Thus we make the
replacement
\begin{equation}\label{S16}
\int_0^t d t'e^{i(\omega_{\vec{k}}-\omega)(t'-t)}\rightarrow
-i\frac{1}{\omega_{\vec{k}}-\omega}
+\pi\delta(\omega_{\vec{k}}-\omega),
\end{equation}
for sufficiently large times. In similar fashion, for large times
the second and third integrals in (\ref{S15}) may be replaced with
\begin{eqnarray}\label{S17}
&&\int_0^t d t'e^{i(\omega_{\vec{k}}+\omega)(t'-t)}\rightarrow
-i\frac{1}{\omega_{\vec{k}}+\omega},\nonumber\\
&&\int_0^t d
t'e^{i(\omega_{\vec{k}})(t'-t)}\rightarrow-i\frac{1}{\omega_{\vec{k}}}.
\end{eqnarray}
Substitution of (\ref{S14}), (\ref{S16}) and (\ref{S17}) in
(\ref{S9}), gives
\begin{eqnarray}\label{S18}
&&a_{\vec{k}\lambda}(t)=a_{\vec{k}\lambda}(0)e^{-i\omega_{\vec{k}}t}-\sqrt{\frac{\alpha^2}{8(2\pi)^3\hbar\omega_{\vec{k}}}}(\vec{k}\times\vec{e}_{\vec{k}\lambda})_x
[\frac{-iS_+(t)}{\omega_{\vec{k}}+\omega}-\frac{iS_-(t)}{\omega_{\vec{k}}-\omega}+\pi\delta(\omega_{\vec{k}}-\omega)]\nonumber\\
&&+\sqrt{\frac{\alpha^2}{8(2\pi)^3\hbar\omega_{\vec{k}}}}(\vec{k}\times\vec{e}_{\vec{k}\lambda})_y[\frac{S_+(t)}{\omega_{\vec{k}}+\omega}-\frac{S_-(t)}{\omega_{\vec{k}}-\omega}-i\pi\delta(\omega_{\vec{k}}-\omega)]\nonumber\\
&&+\sqrt{\frac{\alpha^2}{2(2\pi)^3\hbar\omega_{\vec{k}}}}(\vec{k}\times\vec{e}_{\vec{k}\lambda})_z
\frac{iS_z(t)}{\omega_{\vec{k}}},
\end{eqnarray}
for sufficiently large times.\\
Now substituting $a_{\vec{k}\lambda}(t)$ from (\ref{S18}) in
(\ref{S13.5}) and then taking expectation value of both sides of
(\ref{S13.5}) over the state $
|\psi\rangle=|S\rangle\otimes|0\rangle_F $, where $ |S\rangle $
is an arbitrary state of the spin and $ |0\rangle_F $ is the
vacuum state of the quantum vacuum field, we obtain
\begin{equation}\label{S19}
<\dot{S}_z(t)>=-\beta<S_z(t)>-\beta\frac{\hbar}{2},\hspace{2.00
cm}\beta=\frac{\alpha^2\hbar\omega^3}{6\pi^2c^5},
\end{equation}
where we have used the relations
\begin{eqnarray}\label{S20}
&&\int_0^{2\pi}d\varphi\int_0^\pi\sin\theta
d\theta\sum_{\lambda=1}^2(\hat{k}\times\vec{e}_{\vec{k}\lambda})_x
(\hat{k}\times\vec{e}_{\vec{k}\lambda})_y\nonumber\\
&&=\int_0^{2\pi}d\varphi\int_0^\pi\sin\theta
d\theta\sum_{\lambda=1}^2(\hat{k}\times\vec{e}_{\vec{k}\lambda})_x
(\hat{k}\times\vec{e}_{\vec{k}\lambda})_z\nonumber\\
&&=\int_0^{2\pi}d\varphi\int_0^\pi\sin\theta
d\theta\sum_{\lambda=1}^2(\hat{k}\times\vec{e}_{\vec{k}\lambda})_y
(\hat{k}\times\vec{e}_{\vec{k}\lambda})_z=0,\hspace{1.50
cm}\hat{k}=\frac{\vec{k}}{|\vec{k}|},\nonumber\\
&&\int_0^{2\pi}d\varphi\int_0^\pi\sin\theta
d\theta\sum_{\lambda=1}^2(\hat{k}\times\vec{e}_{\vec{k}\lambda})_x^2=\int_0^{2\pi}d\varphi\int_0^\pi\sin\theta
d\theta\sum_{\lambda=1}^2(\hat{k}\times\vec{e}_{\vec{k}\lambda})_y^2\nonumber\\
&&=\int_0^{2\pi}d\varphi\int_0^\pi\sin\theta
d\theta\sum_{\lambda=1}^2(\hat{k}\times\vec{e}_{\vec{k}\lambda})_z^2=\frac{8\pi}{3}.
\end{eqnarray}
The solution of equation (\ref{S19}) is
\begin{equation}\label{S21}
<S_z(t)>=Ae^{-\beta t}-\frac{\hbar}{2},
\end{equation}
where $A$ is constant that wich can be specified from initial
condition. It is clear from this equation that for any $A$, when $
t $ tends to $ \infty $ we have, $ <S_z(t)> \rightarrow
-\frac{\hbar}{2} $, this means that for any initial state of spin
of particle, the final state tends to $
|-\frac{\hbar}{2}\rangle $ at very large times. For an electron and
$B_L=10^4$ gauss, this time is nearly $ t\simeq\frac{1}{\beta}\simeq 5\times10^6 s $.\\
Also substitution of $ a_{\vec{k}\lambda}(t) $ from (\ref{S18}) in
(\ref{S13}) and then taking expectation value of both sides over $
|\psi\rangle=|S\rangle\otimes|0\rangle_F $, and using (\ref{S20})
\begin{eqnarray}\label{S22}
&&<\dot{S}_+(t)>=i\Omega S_+-\frac{\beta}{2}S_+,\hspace{1.50 cm}\Omega=
\omega+\Delta_1-\Delta_2,\nonumber\\
&&\Delta_1=\frac{\alpha^2\hbar}{12\pi^2c^5}P\int_0^\infty
d\omega_{\vec{k}}\frac{\omega_{\vec{k}}^3}{\omega_{\vec{k}}+\omega},\nonumber\\
&&\Delta_2=\frac{\alpha^2\hbar}{12\pi^2c^5}P\int_0^\infty
d\omega_{\vec{k}}\frac{\omega_{\vec{k}}^3}{\omega_{\vec{k}}-\omega},
\end{eqnarray}
where $P$ denotes the cauchy principal value. Solution of
(\ref{S22}) is
\begin{equation}\label{S23}
<S_+(t)>=Be^{(-\frac{\beta}{2}+i\Omega)t},
\end{equation}
where $B$ is a constant to be determined from initial condition.
In the absence of quantum vacuum, $ \beta=\Delta_1=\Delta_2=0 $
and (\ref{S23}) is reduced to $S_+(t)=Be^{i\omega t}$ which is
the usual spin precession in the presence of the labratory
magnetic field $\vec{B}_L$.

\end{document}